\documentclass[useAMS]{mn2e}
\usepackage{times}
\usepackage{rotating}
\long\def\symbolfootnote[#1]#2{\begingroup%
\def\thefootnote{\fnsymbol{footnote}}\footnote[#1]{#2}\endgroup}
\newcommand{\gae}{\lower 2pt \hbox{$\, \buildrel {\scriptstyle >}\over {\scriptstyle \sim}\,$}}
\newcommand{\lae}{\lower 2pt \hbox{$\, \buildrel {\scriptstyle <}\over {\scriptstyle \sim}\,$}}
\newcommand{\aprop}{\lower 2pt \hbox{$\, \buildrel {\scriptstyle \propto}\over {\scriptstyle \sim}\,$}}

\begin{document}

\title[Radiation from Poynting jet]
{Radiation from a Relativistic Poynting Jet: some general considerations}

\author[Kumar \& Crumley]{Pawan Kumar$^{1}$\thanks
{E-mail: pk@astro.as.utexas.edu, crumleyp@physics.utexas.edu}
and Patrick Crumley$^{1,2}$\footnotemark[1] \\
$^{1}$Department of Astronomy, University of Texas at Austin, Austin, TX 78712, USA\\
$^{2}$Department of Physics, University of Texas at Austin, Austin, TX 78712, USA}

\maketitle

\begin{abstract}
We provide estimates for the flux and maximum frequency of radiation 
produced when the magnetic field in a relativistic, highly magnetized, jet
is dissipated and particles are accelerated using general considerations.
We also provide limits on the jet Lorentz factor and magnetization parameter 
from the observed flux. Furthermore, using the
Lorentz invariance of scalar quantities produced with electromagnetic tensor, 
we provide constraints on particle acceleration, and general features of 
the emergent radiation. We find that the spectrum below the peak softens 
with decreasing frequency. This spectral feature might be one way of 
identifying a magnetic jet.
\end{abstract}


\section{Introduction}

Relativistic jets where the energy is transported outward by Poynting flux 
($\vec E$x$\vec B$) have been invoked for many energetic outflows in 
astrophysical systems such as pulsars, quasars, micro-quasars and 
gamma-ray bursts (GRB).
There is a vast peer reviewed literature on this topic e.g., Michel (1969), Blandford \& Znajek (1977) Blandford \& Payne (1982), Kennel and Coroniti (1884), 
Begelman, et al. (1984), Coroniti (1990), M{\'e}sz{\'a}ros \& Rees\ (1997),
Lyubarsky \& Kirk (2001), de Gouveia dal Pino \& Lazarian (2005), 
Drenkhahn \& Spruit (2002), Lovelace et al. (2002), Kulsrud (2005), Giannios \& Spruit (2006), 
Komissarov et al. (2007), Tchekhovskoy et al. (2008), Metzger et al. (2011),
Cerutti et al. (2012).

The radiation is produced in these systems as a result of
magnetic field dissipation (referred to as {\it reconnection}, 
a generic phrase, which we will be using throughout this article), where
particles are accelerated either via parallel electric field or first order
Fermi process in converging flows, and they then emit photons via the 
synchrotron process. Radiation could also be produced in shocks internal
to the jet or when the jet interacts 
with the surrounding medium via a shock and transfers a fraction of its energy 
to particles in the external medium\footnote{We are considering relativistic 
jets in this work which are Poynting flux dominated such as those that
one encounters in Gamma-ray bursts, disruption of a star by the tidal gravity
of a massive black hole, or AGNs. If the jet energy were to be transported
outward by particles as kinetic energy, then in that case the kinetic energy
could be converted to radiation via internal and external shocks as 
discussed for GRB jets in the works of eg. Meszaros \& Rees (1993),
Rees \& Meszaros (1994), Dermer et al. (1999), Ghisellini \& Celotti (1999),  
Stern \& Poutanen (2004), Beloborodov (2010), Thompson \& Gill (2014); however
the efficiency of converting jet kinetic energy to radiation in internal shocks
is of order only a few percent eg. Kumar (1999). }
We do not consider the latter process in this paper.
   
Magnetic reconnection is a complex and poorly understood process despite the
work of numerous people on this problem for the last more than 50 years, 
e.g. Dungey (1953), Sweet (1958), Parker (1957), 
Petschek (1964), Syrovatskii (1981), Biskamp (1986), Yamada et al. (1997), Kulsrud (1998), Uzdensky \& Kulsrud (2000), Birn et al.(2001), Drake (2006),  Samtaney et al. (2009), Zweibel \& Yamada (2009). Does this mean that we are doomed 
in our effort to understand those astrophysical systems where Poynting jets
play an important role until a predictive theory for reconnections is 
developed? The answer depends on what it is that we want to understand 
about these systems. If we are interested in the general, global, properties 
then the fine details of the reconnection process might not matter. A basic understanding can be obtained from certain 
Lorentz invariant functions of electromagnetic tensor and conservation laws.
The goal of this paper is thus modest, and highly restricted in this sense, 
i.e. to try to provide some constraints on Poynting jet parameters (without
having to rely on a particular reconnection model)
so that magnetic dissipation can explain some broad 
aspects of the data such as the efficiency for
converting magnetic energy to radiation and the general shape of the 
emergent spectrum. In a recent paper Beniamini \& Piran (2014) have
provided constraints on a Poynting jet model for GRBs. Their general
approach and results are very different from the one we pursue here.

In section 2 we provide a few general properties of Poynting jet. We
estimate the maximum energy electrons could achieve in reconnection,  
and the shape of emergent spectrum also in \S2.

\section{Poynting jet: a few general considerations}

\begin{figure}
\begin{centering}
\centerline{\hbox{\includegraphics[width=7.0cm,height=4.0cm,angle=0]{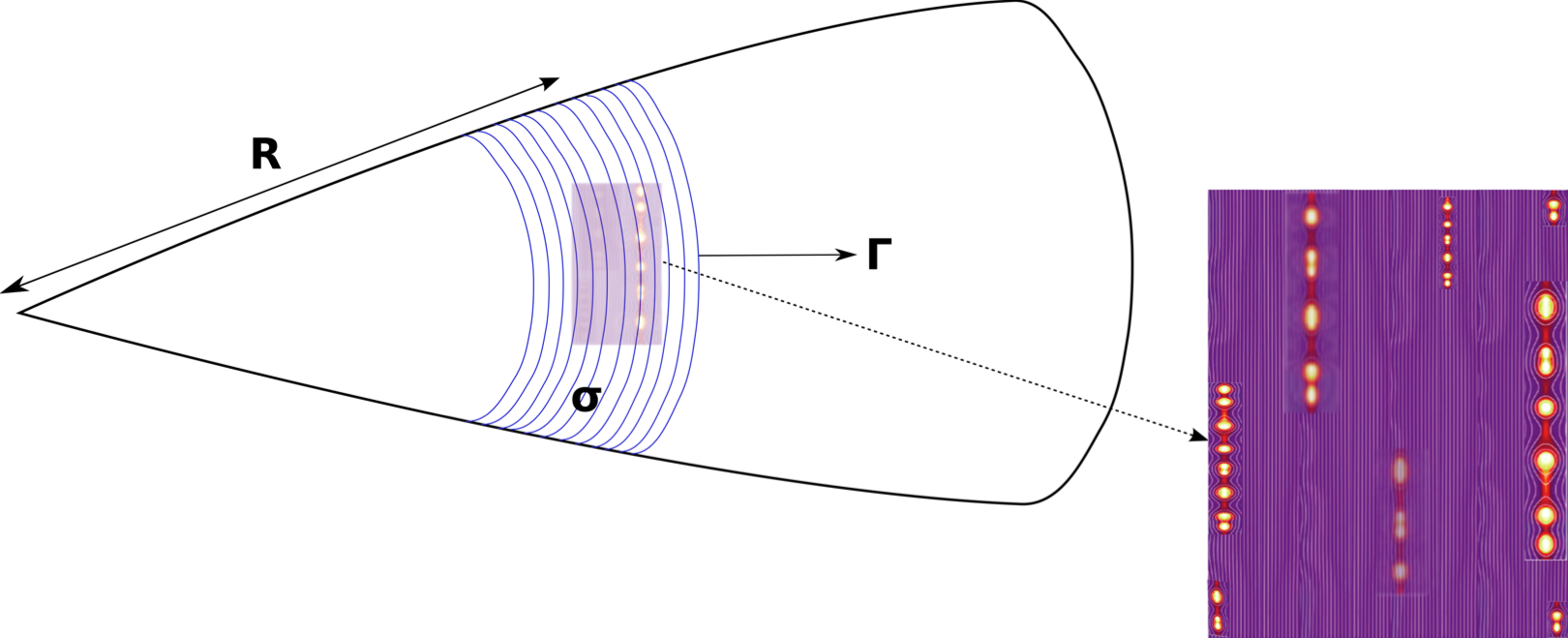}}}
\end{centering}
\caption{A schematic sketch of a Poynting jet and multiple reconnection 
   zones within the causally connected region of comoving size $R/\Gamma$. Each
   reconnection zone has a number of particle-acceleration-sites (PASs), which
   are either regions between bright spots or plasmoids where magnetic field 
   is small and particles are accelerated by electric fields or regions 
   of converging flows where particles are accelerated by the Fermi process. 
   The reconnection zones shown in this figure are artistic renditions of 
   numerical simulations of Hesse \& Zenitani (2007). }
\label{current-sheets} 
\end{figure}

Figure \ref{current-sheets} provides a schematic sketch of the system we are 
considering. The magnetic fields of a relativistic Poynting jet
undergo dissipation at some radius\footnote{The dissipation
of Poynting jet could be spread over a wide range in radius --- $R_1$ to
$R_2$ with $R_2\gg R_1$. We are considering the maximum size of the 
region that is in causal contact at $R_2$, i.e. between $R_2/\Gamma^2$ 
and $R_2$. If this does not capture a good fraction of the energy 
dissipation process then we can add up results from other radii in a trivial
way as processes going on in one region have no effect on another region that
is not in causal contact.}
 $R$ and a number of current sheets form within the causally connected 
region of the jet of comoving size $R/\Gamma$
($R/\Gamma^2$ in lab frame) and efficiently convert the
magnetic energy to particle energy and radiation. Within any current sheet
there are likely to be a number of different regions where particles are
accelerated, and within a space of size $R/\Gamma$ there are obviously 
many more.  These acceleration regions are usually associated with X-points 
--- located in between plasmoids or magnetic islands that form due to 
tearing instability -- where the magnetic field vanishes in absence of a 
guide field and where the electric field can accelerate particles, or 
with converging flows where particle acceleration takes place via first order 
Fermi process.  Regions where particles are accelerated will be referred to as 
PASs (particle acceleration sites). Some general considerations regarding 
particle acceleration in an individual PAS is discussed in \S\ref{particle-acc}.
 The maximum Lorentz factor (LF) of particles 
in PASs is determined by a combination of electric field 
strength\footnote{Particles are also accelerated in converging velocity flows
(first order Fermi process) and stochastic velocity fields (second order Fermi
process), but these are not considered in this work.} and 
radiative losses in addition to energy conservation (\S\ref{particle-acc}).
While outside PASs, particles lose energy to radiation and any acceleration
they experience is negligible. The particle distribution function 
inside PASs is not determined in this paper and is taken to be a 
hard powerlaw function as per numerical simulations (Zenitani \& Hoshino 2001; Jaroschek et al. 2004;  Sironi \& Spitkovsky 2011, 2014; Guo et al. 2014). 
However, the distribution function outside PASs is determined by solving 
an appropriate set of equations (\S\ref{e-dist}). 
Synchrotron radiation emanating from PASs and outside PASs is considered
in \S\ref{spectra}. In \S\ref{number-cs} we provide an estimate for distance 
from the center where Poynting jet is likely dissipated and the number of
current sheets in the causally connected region for an efficient conversion
of magnetic energy to radiation.

Let us consider a Poynting jet with magnetization parameter $\sigma$,
Lorentz factor (LF) $\Gamma$, and isotropic equivalent luminosity $L$. 
The dissipation of magnetic field takes place when the jet is at radius R, 
and that is also roughly the radius where radiation is produced. 
The plasma is sufficiently cold before magnetic reconnections so that 
the thermal pressure of particles can be ignored.

The magnetic field in the jet comoving frame is $B'_0$ --- all physical 
quantities in the jet comoving frame are denoted by a prime and observer 
frame variables are un-primed --- which is related to jet luminosity as
\begin{equation}
      L = B'^2_0 \Gamma^2 R^2 c \quad \Longrightarrow \quad
     B'_0 = { (L/c)^{1/2}\over \Gamma R} = (58\ {\rm G}) {L_{48}^{1/2} \over
          \Gamma_2 R_{15} }
   \label{B-jet}
\end{equation}
provided that $\sigma>1$; we are using the convenient notation $X_n \equiv 
X/10^n$.

We adopt the standard model that charged particles are accelerated in 
reconnection layers where magnetic field dissipation takes place. The 
accelerated electrons with ``thermal'' LF $\gamma'_e$ emit synchrotron
photons of frequency less than or equal to $\nu$ (in the observer frame)
which is given by
\begin{equation}
   \nu \approx {q B'_0 \gamma'^2_e \Gamma\over 2\pi m_e c (1+z)},
   \label{syn-nu}
\end{equation}
where $q$ and $m_e$ are electron charge and mass, and $z$ is the redshift of 
the object. An upper limit to $\gamma'_e$ can be obtained from energy 
conservation, i.e. the energy in accelerated particles cannot exceed the 
energy in magnetic field. This condition gives 
\begin{equation}
    n'_e \gamma'_{max} m_e c^2 \lae {B'^2_0 \over 8\pi} \Longrightarrow 
    \gamma'_{max} \lae (m_p/m_e) \sigma,
    \label{gam-absolute-max}
\end{equation}
where $n'_e$ is electron number density in the jet comoving frame, and
\begin{equation}
     \sigma\equiv B'^2_0/(8 \pi n_e' m_p c^2)
\end{equation}
is jet magnetization parameter.
The reason that equation (\ref{gam-absolute-max}) gives the maximum electron
LF and not the average LF is because electrons accelerated in current sheets
have a power-law distribution function ($d n_e/d\gamma_e\propto \gamma_e^{-p}$)
with $p <2$, and therefore most of the electron ``thermal'' energy is carried
by the highest energy electrons;
 numerical simulations find $p <2$ when the region where the reconnection 
takes place is strongly magnetized, $\sigma\gae$ a few, and when the
reconnection layer is sufficiently large in size 
(e.g. Romanova \& Lovelace 1992; Zenitani \& Hoshino 2001;  
Sironi \& Spitkovsky 2014; Bessho \& Bhattacharjee 2012, Werner et al. 2014, 
Guo et al. 2014). Making use of equations \ref{B-jet} \&
\ref{gam-absolute-max} for magnetic field strength and electron LF, we obtain
an expression for the maximum synchrotron frequency
\begin{equation}
  \nu_{max}^{syn} \sim {q L^{1/2} \sigma^2 (m_p/m_e)^2\over 2\pi m_e c^{3/2} R 
    (1+z)} = (2.2\times10^2{\rm eV}) {\sigma^2 L_{48}^{1/2}\over R_{15}(1+z)}.
\end{equation}
A more accurate estimate for $\nu_{max}^{syn}$ that takes into account
radiative losses are presented in \S\ref{particle-acc}.
The synchrotron photons will be inverse-Compton (IC) scattered to higher
energies by electrons producing these photons, and the maximum IC photon 
energy in observer frame is the smaller of $m_e c^2 \gamma'_e \Gamma/(1+z)$ 
and $\sim \nu \gamma'^2_{max}$.

The specific flux at frequency $\nu$, i.e. flux per unit 
frequency, in the observer frame is 
\begin{equation}
  f_\nu \approx \left[ {q^3 B'_0\Gamma N_e\over m_e c^2}\right] {1+z\over 
       4\pi d_L^2},
   \label{syn-flux}
\end{equation}
where $N_e$ is the total number of electrons (isotropic equivalent) in the
causally connected part of the jet with thermal LF $\ge \gamma_e$, and $d_L$
is the luminosity distance to the source. We can calculate the number of 
electrons needed to produce a given observed flux by combining equations
(\ref{B-jet})--(\ref{syn-flux}):
\begin{equation}
   N_e \approx 1.2\times10^{49} f_{\nu,mJy} L_{48}^{-1/2} R_{15} d_{L,28}^2 
         (1+z)^{-1},
   \label{Ne}
\end{equation}

The optical depth of these electrons to Thomson scattering is,
\begin{equation}
    \tau_T \approx {\sigma_T N_e\over 4\pi R^2} = 8\times10^{-7}\,
       f_{\nu,mJy} L_{48}^{-1/2} R_{15}^{-1} d_{L,28}^2 (1+z)^{-1},
\end{equation}
and their ``thermal'' LF and kinetic energy luminosity they carry are
\begin{equation}
   \gamma'_e \approx \left[ {2\pi m_e \nu R c^{3/2}\over q L^{1/2}(1+z)^{-1}}
          \right]^{1/2} = 4\times10^3 {[R_{15} \nu_{kev}(1+z)]^{1/2}
        \over L_{48}^{1/4}},
\end{equation}

\begin{eqnarray}
    L_e & \approx & {N_e m_e c^3 \gamma_e \Gamma\over (R/\Gamma^2)} \nonumber \\
        & = &(1.2\times10^{42} {\rm erg\, s^{-1}}) {\Gamma^3 f_{\nu,mJy} 
     R_{15}^{1/2} d_{L,28}^2 \nu_{keV}^{1/2} \over L_{48}^{3/4}(1+z)^{1/2}},
\end{eqnarray}
where $\nu_{keV}$ is photon frequency (in units of 1 keV) for which the
observed specific flux is $f_{\nu,mJy}$. Considering that the energy 
carried by electrons cannot exceed the energy in magnetic fields for 
a Poynting jet, we find
\begin{equation}
    L_e/L \lae 1 \quad\Longrightarrow\quad \Gamma \lae 90\; { L_{48}^{7/12} 
     (1+z)^{1/6} \over f_{\nu,mJy}^{1/3} R_{15}^{1/6} d_{L,28}^{2/3} 
     \nu_{keV}^{1/6} }.
\end{equation}
The reason for the approximate inequality sign in the above equation is 
because magnetic fields of a Poynting jet could be compressed by a factor
a few and thus $L_e$ could in principle exceed $L$ by order unity.

If we consider that there are $\eta_p$ protons for every 
electron\footnote{$\eta_p>1$ when only a fraction of electrons in the jet 
are accelerated.} that radiates at frequency $\nu$, then the kinetic 
energy luminosity carried by cold protons is
\begin{equation}
    L_p \approx {N_e \eta_p m_p c^3 \Gamma\over (R/\Gamma^2)} \approx 
        (5\times10^{41} {\rm erg\, s^{-1}}) { \Gamma^3 \eta_p f_{\nu,mJy} 
        d_{L,28}^2 \over L_{48}^{1/2} (1+z) }.
\end{equation}
Therefore, the magnetization parameter for the jet at location where
jet magnetic energy is dissipated and radiation is produced is given by
\begin{equation}
    \sigma(R) \approx {L\over L_p} \approx {2\times10^6 \over \eta_p \Gamma^3}
    f_{\nu,mJy}^{-1} L_{48}^{3/2} d_{L,28}^{-2} (1+z).
\end{equation}
If 10\% of electrons in the jet are accelerated. i.e. $\eta_p = 10$, and
$\Gamma=20$, then $\sigma(R)\approx 25$. And that means that the magnetization
parameter at the let launching site where $\Gamma\sim1$ is
$\sigma_0\approx \Gamma(R)\sigma(R)\sim 500$.

\medskip
\subsection{Particle acceleration in current sheets}
\label{particle-acc}
\medskip

Consider an electron undergoing acceleration in a reconnection
region where the electric field is $\vec E'$, and the magnetic field is
$\vec B'$. In the absence of a guide field, the magnetic field vanishes at 
the X-point, and far away from it
its magnitude is $B_0'$, but otherwise at this stage we place no further
constraint on the electric and magnetic fields. The electron starting from 
some place in the vicinity of the X-point is accelerated, and as it moves 
away it finds the strength 
of the magnetic field increasing. At some point when the magnetic field
becomes sufficiently strong, which will be quantified shortly, the acceleration
ceases if $\vec E'\cdot\vec B'=0$. However, even well before this happens, 
the electron could stop 
accelerating due to radiative losses which will determine its terminal
Lorentz factor. We consider this interplay between acceleration and radiative
losses to determine maximum electron LF.

It is best to view the motion of a particle acted upon by $\vec E'$ and 
$\vec B'$ from a frame where the fields point in the same direction (which 
is always possible except when $|\vec E'| = |\vec B'|$ and the two fields 
are exactly perpendicular to each other). This special frame where
$\vec E''\parallel \vec B''$ will be referred to as the AF frame (Aligned
Fields frame). There are two quadratic Lorentz invariant 
functions of $\vec E'$ and $\vec B'$:  
\begin{eqnarray}
   I_1 & = & -\epsilon_{\alpha\beta\gamma\delta} F^{\alpha\beta} 
       F^{\gamma\delta}/8 =  \vec E'\cdot\vec B' \quad {\it and} \nonumber \\
   I_2 & = & -F^{\alpha\beta} F_{\alpha \beta}/2 = E'^2 - B'^2,
   \label{Lorentz-inv}
\end{eqnarray}
where $F_{\alpha\beta}$ is the electromagnetic tensor (anti-symmetric 2-form).
Since \(\vec{E'}\cdot\vec{B'}\) is Lorentz invariant, if there is 
a non-zero component of magnetic field in the direction of the electric field
in one inertial frame, there will be a non-zero component in all inertial frames. However, 
the component of magnetic field perpendicular to the electric field can be
made to vanish by frame transformation if $E'^2 - B'^2 > 0$, or the
electric field perpendicular to magnetic field can be transformed  away in
an appropriate frame when $E'^2 - B'^2 < 0$. The point is that there exists
an inertial frame (AF) where the transformed electric and magnetic fields
are parallel, and the motion of the electron is the AF frame is as simple as
can be --- the electron momentum parallel to the fields increases linearly 
with time (if the electric field is non-zero in this frame), and the 
perpendicular component of momentum has a constant magnitude (time independent)
and it rotates about the magnetic field at a constant rate. 

\begin{figure}
\begin{centering}
\centerline{\hbox{\includegraphics[width=8.9cm,height=10.7cm,angle=0]{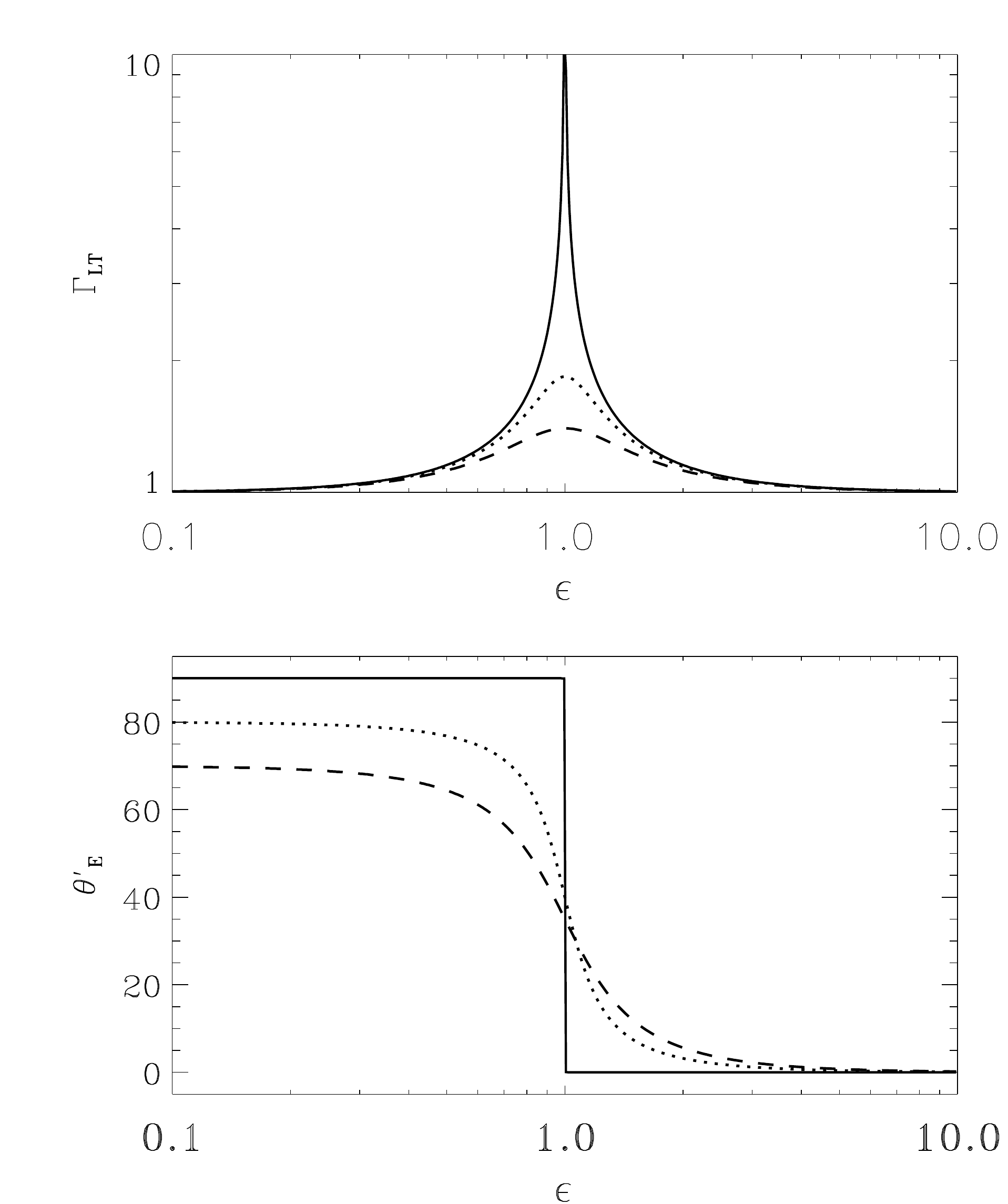}}}
\end{centering}
\caption{The Lorentz factor of the inertial frame (wrt jet rest-frame) in which
    the electric and magnetic fields in some region of current sheet are 
   parallel to each other is shown in the upper panel as a function of 
   $\epsilon\equiv E'/B'$; 
   $\Gamma_{LT} \equiv (1-\beta_{LT}^2)^{-1/2}$, where $\beta_{LT}$ is given 
   by equation (\ref{beta-LT}). The three different curves correspond to three 
   different angles ($\theta'$) between electric and magnetic fields;
   $\theta' = 70^o$ (dashed curve), $80^o$ (dotted line), and $90^o$ 
   (solid line). Note that $\Gamma_{LT}\sim 1$ except when electric field is
   almost exactly perpendicular to the magnetic field and the
   strengths of these fields are about the same. This makes it rather easy 
   to carry out calculations in the AF frame where $\vec E'' \parallel
  \vec B''$ and transform variables back to the jet comoving frame. 
   The lower panel shows the angle (measured in the jet comoving frame in 
   degrees) by which the electric field direction is rotated in the AF frame
   for three different values of $\theta'$ which are same as in the upper panel;
   when the electric field vanishes is the AF frame then the angle is
   the rotation for magnetic field plus $\pi/2$.
   }
\label{LT-gam} 
\end{figure}

The simplest way to get to the AF frame is by a Lorentz boost in the
direction $\vec E'\times\vec B'$; if $\vec E'\times\vec B'=0$, then obviously
no Lorentz transformation is needed as we are already in a frame where the
two fields are either parallel or one of them is zero. A 
straightforward Lorentz transformation algebra shows that the speed of the 
Lorentz boost required so that the fields are parallel in the new frame is
\begin{equation}
    \beta_{LT} = { (1 + \epsilon^2) - \left[ (1-\epsilon^2)^2 + 
    4\epsilon^2\cos^2\theta' \right]^{1/2} \over 2\epsilon \sin\theta' }
    \label{beta-LT}
\end{equation}
where 
\begin{equation}
  \epsilon \equiv {E' \over B'}, \quad \cos\theta' = {\vec E'\cdot\vec B' \over
       |\vec E'| |\vec B'|}.
\end{equation}
Figure \ref{LT-gam} shows the LF of needed boost as a function of 
$\epsilon$ for a few different values of $\theta'$; a simple analytical 
expression for $\epsilon\gg1 $ and $\epsilon\ll1 $ is
\begin{equation}
    \beta_{LT}\approx \min\left\{ \epsilon, \epsilon^{-1}\right\} \sin \theta', 
\end{equation}
which turns out to be exact (as opposed to approximate) for all values 
of $\epsilon$ for the special case of $\theta'=\pi/2$.

The electric field in the new frame follows from the two Lorentz invariant 
quantities mentioned above and is given by
\begin{equation}
  E''^2 = { I_2 + \sqrt{4 I_1^2 + I_2^2}\over 2},
\end{equation}
and the magnetic field is
\begin{equation}
   B'' = {I_1\over E''},
\end{equation}
where $I_1$ and $I_2$ are defined in equation (\ref{Lorentz-inv}). 
The angle between the aligned electro-magnetic field in the AF frame and the
electric field in the jet frame can be easily calculated and is
\begin{equation}
   \cos\theta'_E = {(\epsilon - \beta_{LT}\sin\theta')\over \sqrt{\epsilon^2 + 
  \beta_{LT}^2 - 2\epsilon\beta_{LT}\sin\theta'}},
\end{equation}
The lower panel of Figure \ref{LT-gam} shows $\theta'_E$ as a function of 
$\epsilon$ for a few different values of $\theta'$.

With these results in hand, we are ready to describe particle acceleration in 
a current sheet. Consider a charged particle in the vicinity of the X-point
where $E' \gg B'$. We can transform away the perpendicular component of the 
magnetic field by going to the AF frame, and in this frame the particle 
Lorentz factor $\gamma_e''$ (the double prime emphasizes that we are
in a different inertial frame, and not the jet comoving frame) increases as
$q E'' t''/(m_e c^2)$, which can be rewritten from the point of view of the
jet frame as
\begin{equation}
    \gamma'_e \approx {q \epsilon_0 B'_0 \ell'\over m_e c^2},
\end{equation}
where $\ell'$ is the distance the electron has traveled along the direction
of the electric field from its starting position in the jet comoving frame,
and
\begin{equation}
    \epsilon_0 \equiv E'/B'_0.
\end{equation}

\begin{figure}
\begin{centering}
\centerline{\hbox{\includegraphics[width=9.2cm,angle = 0]{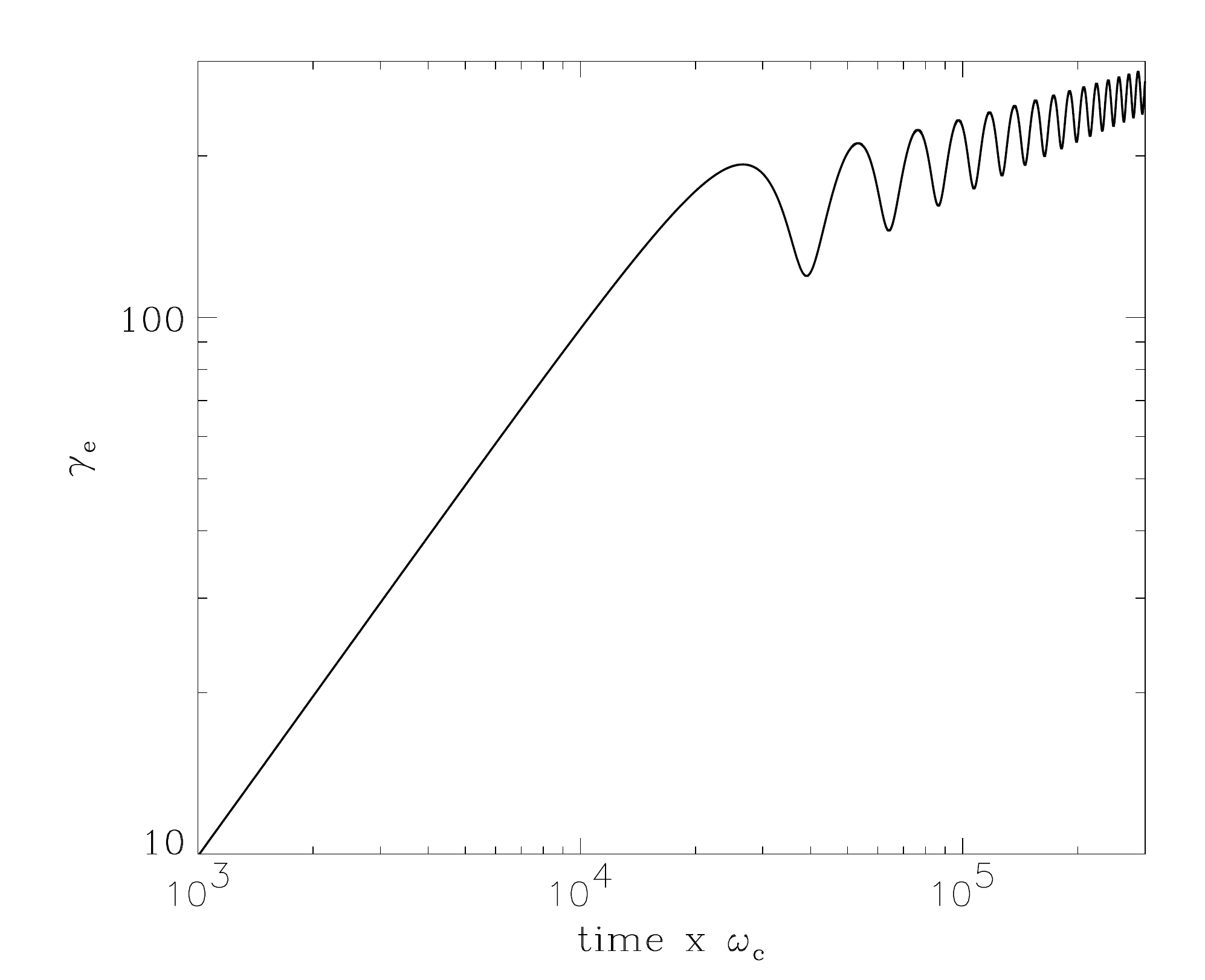}}}
\end{centering}
\caption{Acceleration of an electron in a current sheet is shown as a function 
     of time in unit of $1/\omega_c$, where $\omega_c=q B'_0/(m_e c)$ is Larmor
     frequency. The electric and magnetic field configurations in the 
     current sheet are taken from 
     Larrabee et al (2003), i.e. sheet lies in the x-y plane
     with the electric field pointing in the x-direction and has a constant
     magnitude $E' = \epsilon_0 B'_0$, and the vector
    potential is ${\vec A} = [B'_0(y^2 - z^2)/2\ell'_s] \hat z$; we took 
    $\epsilon_0 = 0.01$ and the length of the ``sheet''  
   $\ell'_s \omega_c/c = 5\times10^5$ for this calculation. The particle
    started out in the $z=0$ plane with initial velocity of zero.
    The LF of the electron ($\gamma'_e$) increases linearly with time as 
    long as it is in the region where $E' > B'$ (which is for about 
    2x10$^4 \omega_c^{-1}$ for the parameters we have chosen for this 
    calculation), and afterward when $B' > E'$ the acceleration ceases and 
    the electron gyrates about the magnetic field. These results are entirely
    consistent with analytical calculations presented in this section.
   }
\label{particle-LF} 
\end{figure}

As the electron travels further and further away from the X-point, it
feels the strength of the magnetic field increase and at some point 
when $B'$ becomes stronger than $E'$ the electron is no longer 
accelerated (unless $\vec E'\cdot\vec B' \not=0$) and its momentum vector 
gyrates about the magnetic field
and the LF oscillations and drifts slowly with time. This 
generic behavior can been in figure (\ref{particle-LF}) where numerical 
result for particle motion in a current sheet is presented.

If the length of the region where $E' > B'$ is $\ell'_E$, then the
maximum LF of electron $\gamma'_{max}\sim q \epsilon_0 B'_0 \ell'_E/(m_e c^2)$;
it should be noted that $\ell'_E \aprop \epsilon_0$ for a 
magnetic field configuration where $B'$ increases linearly with distance
from the X-point, and thus $\gamma'_{max}\aprop \epsilon_0^2$ (see e.g.
Larrabee et al, 2003).
However, two effects can substantially limit electron LF below
this value. One of which is {\it ``global}'' energy conservation, which
provides a limit for $\gamma'_{max}$ as described by equation
(\ref{gam-absolute-max}). And the other is radiative losses  ---
synchrotron and inverse-Compton (IC) for systems of interest to us ---
that could restrict particle LF further. This is discussed below.

Viewed from the AF frame where $\vec E'' \parallel \vec B''$, the electron 
suffers radiative losses due to acceleration along the electric field direction,
gyration about the magnetic field, and inverse-Compton scatterings.
We evaluate each of these to determine the dominant loss mechanism, 
and its effect on $\gamma'_{max}$. The energy loss rate is calculated 
by first assuming that the magnetic 
field lines are parallel, i.e. $\vec B'\cdot\vec\nabla\vec B' = 0$,
 and the electric field is nearly uniform. This
estimate is then improved by relaxing these assumptions and by considering 
the more realistic possibility that magnetic field lines have non-zero
curvature, and that the electric field has spacial fluctuations in the 
acceleration region.

The power radiated due to particle acceleration along the electric field 
can be calculated using the Larmor's formula. The momentum vector of the
electron in the AF frame is nearly parallel to $\vec E''$ since 
it is being accelerated along the electric field and the magnetic field
is parallel to $\vec E''$ in this frame. Therefore, the electric 
field in the instantaneous rest-frame of electron is also $\vec E''$, and
the magnitude of its acceleration in this frame is $q E''/m_e$. It then
follows from Larmor's formula that the power radiated (which is a Lorentz
invariant quantity) is $\sigma_T E''^2 c/4\pi \sim \sigma_T \epsilon_0^2 
B'^2_0 c/4\pi$; from equation 
(\ref{beta-LT}) and Fig. \ref{LT-gam} we know that $\Gamma_{LT} \sim 1$ 
for $\epsilon>2$, and hence $E''\sim E' = \epsilon_0 B'_0$. This 
rate of loss of energy is independent of electron LF,
 and so the maximum electron LF in this case is bounded only by the
size of the acceleration region.

The synchrotron
loss rate due to electron gyration about the magnetic field is $\sigma_T B''^2
\gamma''^2_\perp \beta''^2_\perp c/(6\pi)$; where
$m_e c \gamma''_\perp \beta''_\perp$ is the 4-momentum of
the electron perpendicular to the magnetic field.
In the region where particles are undergoing acceleration,
the magnetic field in the AF frame ($B''$) either vanishes (if 
$\vec E'\cdot\vec B'=0$) or is parallel to $\vec E'$, and in either case
the value of $\gamma''_\perp\beta''_\perp$ does not change with time
even as the electron continues to accelerate.
Thus, the synchrotron loss rate (like the loss rate due to acceleration along
the electric field) is nearly independent of electron momentum, which 
continues to increase linearly with time along $\vec E'$ while the electron 
is in the acceleration region.

For realistic astrophysical systems we don't expect the magnetic and
electric field lines to be perfectly straight in the acceleration
region. The curvature of field lines and the variation of 
$E'/B'$ with distance from the X-point causes the direction of $\vec E''$ 
to change (see Fig. \ref{LT-gam} for the dependence of $\theta'_E$ on $E'/B'$),
and therefore particle momentum vector is also rotated. Due to these
effects $\gamma''_\perp$ is no longer independent of time, and in fact
 even a modest curvature in $\vec E'$ would lead to $\gamma''_\perp\sim
\gamma''_\parallel$. In this case the synchrotron loss estimated above
increases by a factor $\gamma'^2$ (the loss due to acceleration
along $\vec E'$ increases by a similar factor) and is given by
\begin{equation}
   {d m_e c^2 \gamma'^2\over dt'} \sim \sigma_T (\epsilon^2_0 + \sin^2\theta'_g)
    B_0'^2 \gamma'^2 c/6\pi \sim \sigma_T B_0'^2 \gamma'^2 c/6\pi
\end{equation}
where $B'_0\sin\theta'_g$ is the strength of the guide field. From here
on we assume that the guide field is not much smaller than $B'_0$ and 
therefore particle acceleration is dominated by electric field
parallel to the magnetic field. 

The inverse-Compton loss rate is proportional to the energy density
of photons, which is closely related to magnetic field dissipation.
Photons are produced via the synchrotron process in acceleration
regions and also outside it. Assuming that
a fraction $\psi_B$ of magnetic field energy in a causally connected 
region of size $R/\Gamma$ is dissipated in a dynamical time and converted 
to radiation, the photon energy density in the comoving jet frame 
is\footnote{Photons
produced by the dissipation of magnetic field in a region of size $R/\Gamma$
in the comoving frame travel a distance in a dynamical time which is also 
$R/\Gamma$, and hence all the radiative energy is confined
to a volume $\sim R^3/\Gamma^3$.}  $\psi_B B'^2_0/8\pi$,
and therefore the IC loss rate is
\begin{equation}
   {d m_e c^2 \gamma'^2\over dt'} = \sigma_T \psi_B B'^2_0 \gamma'^2 c/6\pi.
\end{equation}

Equating the rate of energy gain for an electron as it is accelerated along
the electric field with the rate of radiative loses we arrive at the following
equation for the maximum value for LF
\begin{equation}
     q \epsilon_0 B'_0 c \approx {\sigma_T B'^2_0 \gamma'^2_{max} c 
     (1 + \psi_B)\over 6\pi}.
\end{equation}
Or
\begin{equation}
   \gamma'_{max} \approx 1.5\times10^7 \epsilon_0^{1/2} (1 + 
    \psi_B)^{-1/2} \Gamma_2^{1/2} L_{48}^{-1/4} R_{15}^{1/2},
    \label{gam-max}
\end{equation}
where we made use of equation (\ref{B-jet}) to substitute for $B_0'$, and
we are considering the case where $\vec E'\cdot\vec B'\not=0$.
The maximum electron LF is the smaller of values given in equations
(\ref{gam-absolute-max}) and (\ref{gam-max}).

The distance an electron travels to get accelerated to $\gamma_{max}$ is
\begin{eqnarray}
    \ell_a' & \approx & {\gamma'_{max} m_e c^2\over q\epsilon_0 B'_0}\nonumber\\
            & \approx &  \min \left\{ \begin{array}{l}
   \hskip -5 pt ( 4.7\times10^8{\rm cm})\, {\Gamma_2^{3/2} L_{48}^{-3/4}
          R_{15}^{3/2}\over \epsilon_0^{1/2} (1 + \psi_B)^{1/2}}, \\ \\
   \hskip -5 pt  (5.2\times10^4{\rm cm})\, \epsilon_0^{-1}\sigma L_{48}^{-1/2}
          \Gamma_2 R_{15}       \\
\end{array} \hskip -4 pt \right\}
   \label{la}
\end{eqnarray}
The synchrotron photon energy corresponding to $\gamma'_{max}$ is 
\begin{equation}
    \nu_{max} \sim 
           \min \left\{ \begin{array}{l} 
\hskip -5 pt (150\,{\rm MeV})\Gamma{\epsilon_0
     (1 + \psi_B)^{-1}}, \\ \\
   \hskip -5 pt (200\, {\rm eV})\; \sigma^2 L_{48}^{1/2}  R_{15}^{-1}      \\
    \end{array} \hskip -4 pt \right\}
\end{equation}

The minimum electron LF can be obtained by taking the length of the 
acceleration region to be no less than proton Larmor radius. This gives
$\gamma'_{min} \sim \epsilon_0^2 (m_p/m_e)$.

\subsection{Electron distribution function}
\label{e-dist}

Simulations of particle acceleration in a reconnection layer show that
the energy distribution is a hard powerlaw function below $\gamma'_{max}$
and exponentially cut-off above it, i.e. $dn'_e/d\gamma'_e \propto 
\gamma'^{-p_0}_e$, for $\gamma'_e < \gamma'_{max}$ with $p_0 < 2$. 

When we add up particle distribution functions in all PASs within the 
causally connected region of the jet at the radius where a good fraction 
of magnetic energy in the jet is dissipated, the resulting distribution is 
\begin{equation}
   {d n'_e\over d\gamma'_e} \propto \gamma'^{-p}_e  \quad\quad\quad\quad {\rm for}\quad
         \gamma'_{min} \lae \gamma'_e \lae \gamma'_{p}.
\end{equation}
The value of $p$ depends on how many electrons pass through PASs which can
accelerate them to LF $\gamma'_e$; if the number of PASs 
increases rapidly with decreasing $\gamma'_{max}$ then $p>p_0$.
The electron distribution below $\gamma'_{min}$ is either 
cutoff or drops off such that the total number of electrons with $\gamma'_e <
\gamma'_{min}$ is small and can be ignored. The distribution above $\gamma'_p$
also falls off more rapidly than $\gamma'^{-2}_e$.

Let us assume that electrons spend an average of $t'_{cs}$ time in an
acceleration region which is larger than $\ell'_a/c$; the average is 
taken over all particle acceleration sites or PASs in causally connected 
part of the jet.  Furthermore, the total number of 
electrons injected into these acceleration regions, in the causally
connected part of the jet, per unit time is
$\dot N'_{e,cs}$. The average rate at which particles exit PASs should also 
be $\dot N'_{e,cs}$. Particles outside PASs cool down radiatively and 
therefore the particle distribution outside is much steeper. We calculate 
this distribution, and estimate the synchrotron flux from electrons inside
and outside PASs.

{If the average time spent by electrons outside PASs is
$t'_{dz}$, then in that time electrons cool down to LF }
\begin{equation}
    \gamma'_c \approx {6\pi m_e c\over \sigma_T B'^2_0 t'_{dz} } \approx
      {6\pi m_e c^3 \Gamma^3 R \xi \over \sigma_T L} \approx \xi \Gamma_1^3
    R_{15} L_{48}^{-1},
    \label{gam-c}
\end{equation}
where
\begin{equation}
    \xi \equiv {R \over c\Gamma t'_{dz}},
\end{equation}
is the ratio of dynamical time in jet comoving frame and $t'_{dz}$.
 
The electron distribution outside PASs is obtained by solving 
\begin{equation}
    {\partial (dN'_{e,dz}/d\gamma'_e)\over \partial t'} + {\partial \dot
    \gamma'_e (dN'_{e,dz}/d\gamma'_e)\over \partial \gamma'_e} = S'(\gamma'_e),
   \label{ne-dist}
\end{equation}
where the source function is
\begin{equation}
     S'(\gamma'_e) \approx {(p-1) \dot N'_{e,cs} \over \gamma'_{min} }
             \left( {\gamma'_e\over \gamma'_{min} }\right)^{-p}  \quad
             {\rm for} \;  \gamma'_{min} \le \gamma'_e
                             \le \gamma'_p
   \label{source-fn1}
\end{equation}
$p > 1$, $\dot N'_{e,cs}$ is the rate at which electrons with Lorentz factors 
$\geq \gamma'_{min}$ leave acceleration regions and enter the surrounding 
medium, and 
\begin{equation}
   \dot\gamma'_e = -{\sigma_T B'^2_0 \gamma'^2_e \over 6\pi m_e c}.
\end{equation}
A quasi-steady state solution is reasonable to consider when the time it 
takes for a typical PAS in the jet to form and disappears is much shorter 
than the dynamical time, and there are many PASs in the causally connected 
region of the jet that contribute to particle acceleration and radiation; 
the average of all these PASs can be taken to be roughly constant for 
about a dynamical time. The solution of 
equation (\ref{ne-dist}) for $p\ge1$, in quasi-steady state, is
easy to obtain and for $\gamma'_c < \gamma'_{min}$ is given by
\begin{equation}
    {d N'_{e,dz} \over d\gamma'_e} \approx {t'_{dz} \dot N'_{e,cs}
       \over \gamma'_c} 
       \left\{ \begin{array}{ll}
        \hskip -7 pt {\gamma'^2_c \gamma'^{-p-1}_e \over \gamma'^{-p+1}_{min} }
     \quad\quad  &  \gamma'_{min} \le \gamma'_e \le \gamma'_p \\ \\
        \hskip -7 pt \left( {\gamma'_e \over \gamma'_c} \right)^{-2} &
                      \gamma'_c \le \gamma'_e \le \gamma'_{min}  \\
      \end{array} \right.
     \label{ne-dist-sol}
\end{equation}
The above derivation assumes $\gamma'_p\gg\gamma'_c$, which should be a good 
approximation considering that $\gamma'_p\sim \gamma'_{max} \sim 10^7$ 
(eq. \ref{gam-max}) and $\gamma'_c\sim 1$ (eq. \ref{gam-c}). The
distribution is effectively cutoff above $\gamma'_p$ outside the PASs since 
electrons of this high energy cool efficiently and their LF drops below 
$\gamma'_p$ quickly. The distribution function for the case where 
$\gamma'_p > \gamma'_c > \gamma'_{min}$ is
\begin{equation}
    {d N'_{e,dz} \over d\gamma'_e} \approx t'_{dz} \dot N'_{e,cs}
       \left\{ \begin{array}{ll}
        \hskip -7 pt {\gamma'_c \gamma'^{-p-1}_e \over \gamma'^{-p+1}_{min} }
      \quad\quad\quad &  \gamma'_{c} \le \gamma'_e \le \gamma'_p \\ \\
        \hskip -7 pt {\gamma'^{-p}_e \over \gamma'^{-p+1}_{min} } &
                      \gamma'_{min} \le \gamma'_e \le \gamma'_c  \\
      \end{array} \right.
     \label{ne-dist-sol1}
\end{equation}
For $p<1$, the source function is
\begin{equation}
     S'(\gamma'_e) \sim {(1-p) \dot N'_{e,cs} \over \gamma'_p }
             \left( {\gamma'_e\over \gamma'_p }\right)^{-p}  \quad
             {\rm for} \;  \gamma'_{min} \le \gamma'_e
                             \le \gamma'_p
   \label{source-fn2}
\end{equation}
and therefore most of the electrons are at $\gamma'_e\sim \gamma'_p$.
The solution of equation (\ref{ne-dist}) using the above source function
for $\gamma'_c < \gamma'_{min}$ is
\begin{equation}
    {d N'_{e,dz} \over d\gamma'_e} \approx {t'_{dz} \dot N'_{e,cs}\over 
         \gamma'_c} \left( {\gamma'_e \over \gamma'_c} \right)^{-2}  
           \quad\quad\quad{\rm for}\quad \gamma'_c < \gamma'_e \lae \gamma'_p.
     \label{ne-dist-sol2}
\end{equation}
And the distribution function when $\gamma'_p > \gamma'_c > \gamma'_{min}$ 
is given by
\begin{equation}
    {d N'_{e,dz} \over d\gamma'_e} \approx t'_{dz} \dot N'_{e,cs}
       \left\{ \begin{array}{ll}
        \hskip -7 pt {\gamma'_c \over \gamma'^2_e}
      \quad\quad\quad\quad\quad &  \gamma'_{c} \ll \gamma'_e \le \gamma'_p \\ \\
        \hskip -7 pt {1\over \gamma'_p} \left( {\gamma'_e \over \gamma'_p}
             \right)^{-p} & \gamma'_{min} \le \gamma'_e \ll \gamma'_c  \\
      \end{array} \right.
     \label{ne-dist-sol3}
\end{equation}
The apparent discontinuity of the distribution function in 
equation (\ref{ne-dist-sol3}) at $\gamma'_e=\gamma'_c$ is because the 
two branches of solutions are inaccurate as $\gamma'_e$ approaches
$\gamma'_c$. However, a steep drop off of the distribution function just 
below $\gamma'_c$ is physical. This is due to 
the fact that electrons with LF $\sim \gamma'_p$ (which are a majority of the 
electrons entering the medium in between PASs when $p<1$) radiatively cool down
to LF $\sim \gamma'_c$ in the available time $t'_{dz}$, and hence there 
is an accumulation of electrons in the neighborhood of $\gamma'_c$ and that
is responsible for a drop in the distribution function just below this LF.

\subsection{Synchrotron and IC spectra}
\label{spectra}

{Electrons inside PASs gain energy due to 
acceleration along electric fields or as a result of Fermi mechanism 
operating in a converging flow field. The balance between energy loss 
and gain determines the terminal Lorentz factor for particles.  
Moreover, the particle distribution function, and index $p$, are also 
determined by the acceleration and radiative loss processes, 
and thus the synchrotron spectrum due to radiation from electrons inside PASs
is}
$f_\nu \propto \nu^{-(p-1)/2}$ for $\nu \lae \nu_p$; where 
\begin{equation}
 \nu_p \approx {q B'_0\gamma'^2_p \Gamma\over 
    2\pi m_e c(1+z)} \approx {q L^{1/2} 
   \gamma'^2_p \over 2\pi m_e c^{3/2} R(1+z)}
\end{equation}
is synchrotron frequency in the observer frame corresponding to electron 
LF $\gamma'_p$; the electron distribution function starts to fall off faster
than $\gamma'^{-p}_e$ for $\gamma'_e > \gamma'_p$. 
The specific flux (flux per unit frequency) at $\nu_p$ due to PAS electrons is
(e.g. Rybicki and Lightman, 1979)
\begin{eqnarray}
 f_{cs}(\nu_p)&\sim & {q^3 B'_0\Gamma
     N_e(\gae\gamma'_p)\over m_e c^2}{1+z\over 4\pi d_L^2}\quad\quad\nonumber \\
      &\sim & {q^3 L^{1/2} \dot N'_{e,cs}
       t'_{cs} \over m_e c^{5/2} R} {1+z\over 4\pi d_L^2}  
        \left\{ \begin{array}{ll}
       \hskip -7pt 1          &  p < 1  \\ \\
       \hskip -7pt \left[{\gamma'_p\over\gamma'_{min}}\right]^{1-p}  &  p > 1 \\
       \end{array}   \right.  \quad
     \label{flux-nup1}
\end{eqnarray}
where $N_e(\gae\gamma'_p)$ is the total number of electrons inside PASs
with LF $\gae\gamma'_p$ (which is $t'_{cs}$ times the integral of the 
source function given in equations \ref{source-fn1} \& \ref{source-fn2}), 
$t'_{cs}$ is the average time electrons spend in acceleration regions,
and $d_L$ is the luminosity distance of the source at redshift $z$.

The synchrotron spectrum due to electrons outside PASs is 
either $f_\nu \propto \nu^{-p/2}$, $\nu^{-1/2}$ or $\nu^{-(p-1)/2}$ depending 
on whether $p$ is larger or smaller than 1, and the ordering of $\nu$ and
synchrotron characteristic frequencies. 

The synchrotron flux at $\nu_p$ due to electrons outside PASs
can be estimated using the distribution function
calculated in the previous subsection (eqs. \ref{ne-dist-sol}, 
\ref{ne-dist-sol1}, \ref{ne-dist-sol2}, \ref{ne-dist-sol3}). 
For the case we are considering where the guide field strength is of 
order $B'_0$, the magnetic fields outside and inside PAS are of similar
strength, and in that case the flux at $\nu_p$ due to electrons outside 
PASs is
\begin{equation}
  f_{dz}(\nu_p) \sim 
       f_{cs}(\nu_p) \left[ { t'_{cool}(\gamma'_p) \over t'_{cs} }\right],
    \sim f_{cs}(\nu_p) \left[ {t'_{dz} \gamma'_c\over t'_{cs} \gamma'_p}\right] 
\end{equation}
where $f_{cs}(\nu_p)$ is synchrotron flux at $\nu_p$ due to electrons inside
PASs (see eq. \ref{flux-nup1}), and 
\begin{equation}
    t'_{cool}(\gamma'_p) = {6\pi m_e c\over \sigma_T B'^2_0 \gamma'_p} 
        = ({\rm 2.3\, s}) L_{48}^{-1} \gamma'^{-1}_{p,5} \Gamma_2^2 R_{15}^2
     \label{tcool}
\end{equation}
is synchrotron cooling time for an electron of LF $\gamma'_p$ outside PASs.
Therefore, the ratio of synchrotron flux at $\nu_p$ due to electrons inside 
and outside PASs is
\begin{equation}
    {\cal R}_p \equiv {f_{cs}(\nu_p)\over f_{dz}(\nu_p) } \sim 
     {t'_{cs}\over t'_{cool}(\gamma'_p)}.
\end{equation}
At a frequency $\nu$ between $\nu_{min}$ and $\nu_p$ (assuming that 
$\nu_{min}>\nu_c$), the ratio of the flux from the two regions is\footnote{
$\nu_{min}$ and $\nu_c$ are synchrotron frequencies in the observer frame for 
electrons of Lorentz factors $\gamma'_{min}$ and $\gamma'_c$ respectively
in a magnetic field of strength $B'_0$; $\gamma'_c$ is given by
eq. \ref{gam-c}, and $\gamma'_{min}$ is the electron LF below which the average
distribution function inside PASs either drops off or rises less rapidly 
than $\gamma'^{-p}_e$.}
 
\begin{equation}
   {f_{cs}(\nu)\over f_{dz}(\nu) }\approx {\cal R}_p \left\{ \begin{array}{ll}
   \hskip -5 pt (\nu/\nu_p)^{(2-p)/2}  \quad\quad\quad\quad   &  1/3<p<1  \\ \\
   \hskip -5 pt (\nu/\nu_p)^{1/2}               &    \quad\quad\;\;\;  p> 1  \\
\end{array} \right.
\label{flux-rat1}
\end{equation}
And the flux ratio at a frequency such that $\nu_c < \nu < \nu_{min}$ is
\begin{equation}
   {f_{cs}(\nu)\over f_{dz}(\nu) }\approx {\cal R}_p \left\{ \begin{array}{ll}
   \hskip -7 pt \left( {\nu\over \nu_{min}} \right)^{5/6} \left[ 
      {\nu_{min}\over \nu_p} \right]^{(2-p)/2}   &  1/3<p<1  \\ \\
   \hskip -7 pt \left( {\nu\over \nu_{min}} \right)^{5/6} 
           \left[ {\nu_{min}\over \nu_p}\right]^{1/2}  &   p> 1  \\
\end{array} \right.
\label{flux-rat2}
\end{equation}

\begin{figure}
\begin{centering}
\centerline{\hbox{\includegraphics[width=8.9cm,height=10.7cm,angle=0]{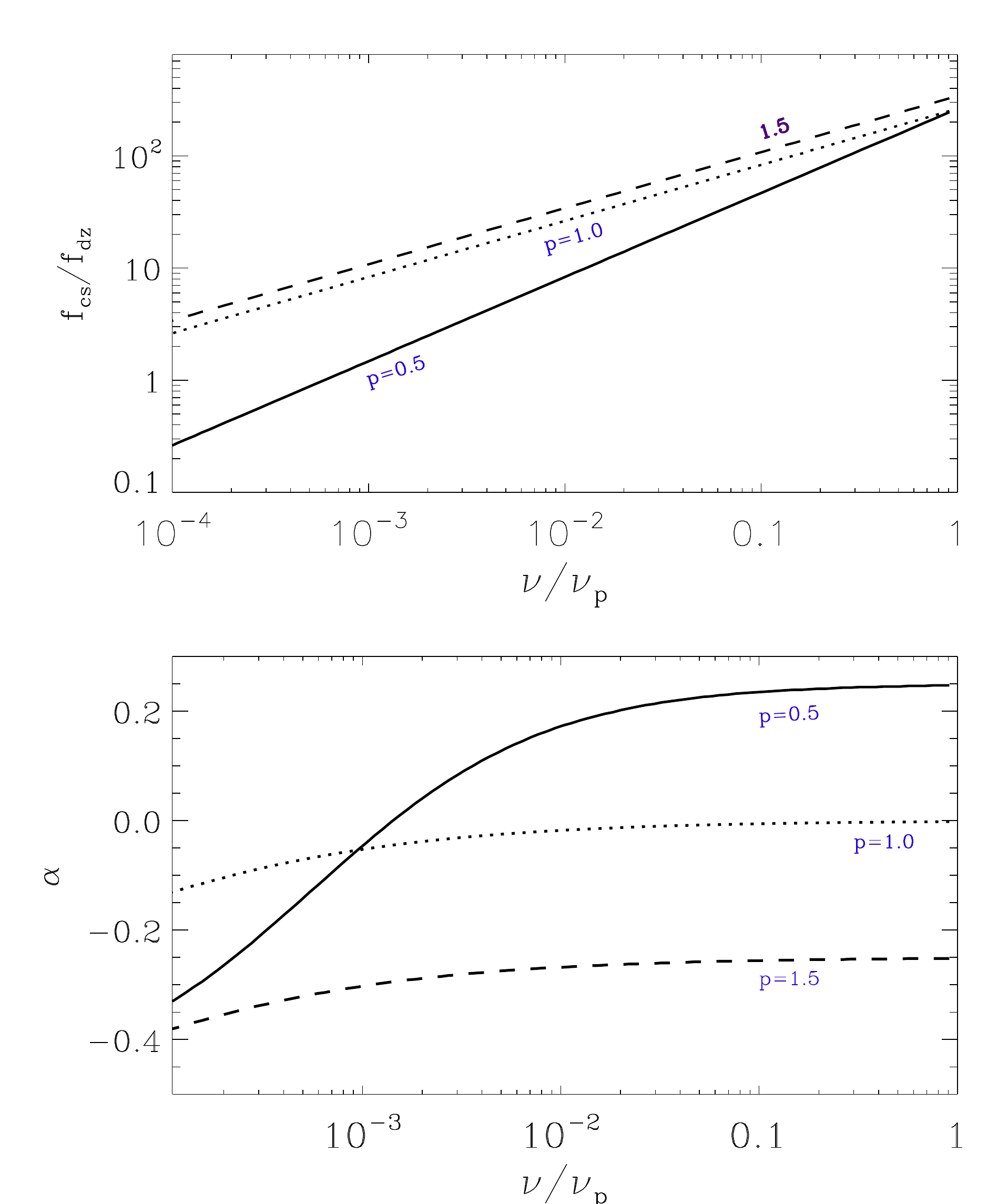}}}
\end{centering}
\caption{The upper panel shows the ratio of synchrotron flux from electrons in
    acceleration regions (PASs) and electrons in regions outside; three 
    different lines correspond to three different values of $p$: 0.5 
    (solid line), 1.0 (dotted line) and 1.5 (dashed line). The parameters 
    for these calculations are:  $L=10^{48}$erg/s, $\Gamma=10$, $\sigma=10^2$, 
    $R=10^{15}$cm, $\epsilon_0=0.1$, $\psi_B=0.1$, $\gamma'_p =\gamma'_{max}$,
    $\gamma'_{min} = \gamma'_{max}/10^2$, and $t'_{cs} = 10^{-3}\times 
    R/(c\Gamma)$. The lower panel shows the spectral index for the
    observed flux, i.e. $\alpha = d\ln(f_{cs} + f_{dz})/d\ln\nu$ for the 
    same three values of $p$. Note that for the parameters chosen for these 
    calculations the observed flux is dominated by electrons radiating in 
    PASs for $\nu \gae 10^{-3} \nu_p$ and therefore the spectral index in this
    frequency range is $\alpha\approx -(p-1)/2$, i.e. the spectrum is hard. }
\label{flux-alpha} 
\end{figure}

The observed spectrum is a superposition of synchrotron radiation from 
electrons in PASs and electrons outside acceleration regions. We have 
provided all the relevant equations to determine specific flux at an
arbitrary frequency from these two sources. 
We note that if the observed flux between $\nu_{min}$ and $\nu_p$ is 
dominated by synchrotron radiation within PASs then the spectrum would be
$f_\nu\propto \nu^{-(p-1)/2}$, which is harder by $\nu^{1/2}$ than the 
case where the flux comes mostly from electrons in the medium between PASs.
Figure \ref{flux-alpha} shows the relative contributions of the two sources
(upper panel), and the spectral index of the observed flux (lower panel), 
as a function of frequency. The figure clearly shows that the flux at 
$\nu_p$ is dominated by electrons in PASs and thus the spectral index is 
harder. Electrons outside PASs make the dominant contribution to the observed 
flux at sufficiently low frequencies (below $\nu_p/10^2$ for the parameters
considered in Fig. \ref{flux-alpha}), and therefore the spectrum is 
softer. This behavior --- softening of spectrum with decreasing frequency 
--- is opposite to what we observe in the synchrotron radiation from particles 
that are accelerated in shocks. This spectral feature may be a way to 
determine if magnetic or shock dissipation of jet energy is responsible 
for the observed radiation. 

In practice, the spectral softening before the peak could be produced in 
a baryonic jet if there are multiple emission processes at work, for
instance synchrotron radiation with a thermal component or synchrotron 
self-Compton (SSC). For a synchrotron + thermal component, the 
expected softening at low energies is very large which cannot be
confused with the softening we expect for reconnection described
above. It is more difficult to tell apart a kinetic jet with SSC
radiation mechanism and a magnetic jet where synchrotron dominates.
The SSC should produce two 
distinct peaks in the spectrum, one for synchrotron and one for 
inverse Compton, and in this case the radiation at lower energies 
dominated by the synchrotron process can be softer. The spectrum
for the magnetic model on the other hand has just one peak, and 
therefore in principle it can be distinguished from the SSC model.
However, in practice, the first peak for the SSC model may be at 
a frequency that is below the observing band, and that would
make the task of identifying a Poynting jet more difficult. There are 
examples of low energy spectral softening in astrophysical objects, 
e.g., GRB 090926A (Ackermann et al. 2011), but it is difficult to say 
whether or not it is due to multiple components to the spectrum or from the 
spectral feature described in this paper. We looked for the spectral 
feature we predict for a magnetic reconnection model in solar flares where
magnetic dissipation is widely believed be at work (e.g. Lin et. al, 2003).
Unfortunately, solar flares have a large thermal component, 
so the presence of a soft spectral feature at low frequencies 
is difficult to discern in these transients.

The ratio of the luminosity in synchrotron and IC radiations is equal
to the ratio of energy densities in magnetic field and photons. Since
only a small fraction of energy in magnetic fields in dissipated 
in current sheets\footnote{In general, it is highly unlikely for magnetic
fields on the opposite sides of currents sheets to be 
exactly anti-parallel, and that limits the efficiency for converting 
magnetic energy to particle energy and radiation.}, it is expected 
that the IC luminosity of a Poynting jet would be smaller than 
the synchrotron luminosity.
The IC spectrum, or to be precise synchrotron-self-Compton
spectrum, is straightforward to calculate using the results for 
particle distribution and synchrotron spectrum described above.

\subsection{Constraint on the number of current sheets in the jet}
\label{number-cs}

The radius interval where conversion of magnetic energy to thermal energy 
for a Poynting jet takes place depends on the magnetic field 
configuration and instabilities that develop in the jet. These are very 
difficult to calculate with any confidence.
However, some general considerations described in this subsection 
provide broad guidance which can be used to constrain the dissipation 
radius and the number of current sheets in the jet.

Consider a short segment of the jet that was launched at radius $R_0$. 
The dissipation of magnetic energy
in this segment could take place anywhere between $R_0$ and $\sim R_d$
(the deceleration radius of the jet), 
either gradually over this long distance interval or suddenly 
within a short distance. 
Once the reconnection gets started at one location in the jet, it could 
trigger magnetic field dissipation at other sites
--- possibly as a result of plasma outflow from this region or 
magnetic field reconfiguration propagating at Alf\'ven speed and 
triggering reconnection at other sites --- and these {\it secondary} 
reconnection sites lie in a region of the jet that is in causal contact 
with the current sheet triggering these events.
It is unlikely that most of the energy of this
segment of the jet under consideration will be dissipated at a radius 
smaller than $(\delta t)c \Gamma^2$
because of causality considerations (provided of course that different parts
of this segment of the jet don't independently develop instability and/or
reconnection centers). We can describe the dissipation with radius 
as a monotonically increasing function of radius, $\zeta_B(R)$; 
$\zeta_B$ is the fraction of the magnetic energy in the segment
that is dissipated or converted to bulk kinetic
energy of the jet.

For reconnections in a jet consisting of
stripped magnetic wind geometry (magnetic fields reversing direction 
over distance of $r_0$ in the lab frame), $\zeta_B\propto R^{1/3}$ (e.g. 
Drenkhahn and Spruit, 2002; Kumar \& Zhang, 2014), and the process
is completed at a radius $R_c \sim r_0 \Gamma^2/\epsilon_0$; 
where $\epsilon_0$ as defined in
\S\ref{particle-acc} is the ratio of electric and magnetic fields and 
$v'_p\sim\epsilon_0c$ is the speed for plasma flow into current sheets.
So, although, the
magnetic field dissipation process in this case is slow and extends over 
a large distance interval of $R_0$ --- $r_0 \Gamma^2/\epsilon_0\gg R_0$, 
roughly half of the magnetic energy is in fact dissipated within a factor 
of a few of $r_0 \Gamma^2/\epsilon_0$. This follows from causality, i.e. the
size of the region where magnetic field is dissipated cannot increase at
a speed faster than light, and hence the radial width of region where
field has been dissipated grows proportional to $R/\Gamma$ 
(in jet comoving frame) and that is the reason that a good fraction of
magnetic energy dissipation occurs within a factor a few of the terminal
radius where the dissipation process is completed. 
This property is likely to be generic, and independent of magnetic field
geometry and reconnection model.

Current sheets are likely to form and disappear on a time short compared
with the dynamical time ($R/(c\Gamma)$ in jet comoving frame). 
We envision that there are an average $\aleph_s$ PASs present in the
region at any given time, for a time duration $\sim R/(c\Gamma)$, and the 
average length of these current sheets in jet comoving frame is $\ell'_s$. 

The total volume of plasma in the jet in the causally connected region
at $R$ is 
\begin{equation}
    {\cal V}'_c \sim (R/\Gamma)^3, 
\end{equation}
provided that the jet opening angle is $>\theta_j$. Since plasma flows
into current sheets at speed $v'_p\sim \epsilon_0 c$, the total volume of 
plasma passing through current sheets in a dynamical time is 
\begin{equation}
    {\cal V}'_{plasma,cs} \sim \aleph_s{\ell'_s}^2 \epsilon_0 (R/\Gamma).
\end{equation}
If the fraction of the magnetic energy in the jet dissipated in this 
region is $\zeta_B(R)$, then that means that the total volume of plasma passing
through current sheets in volume ${\cal V}'_c$ should be $\zeta_B {\cal V}'_c$.
Thus, we obtain the number of PASs in the region to be
\begin{equation}
   \aleph_s \sim [\zeta_B(R)/\epsilon_0] \left\{R/(\Gamma\ell'_s)\right\}^2.
   \label{N-sheet}
\end{equation}

A lower limit for $\aleph_s$ can be obtained by substituting $\ell'_s \sim
R/\Gamma$ in equation (\ref{N-sheet}), which gives
\begin{equation}
    \aleph_s \gae \zeta_B(R)/\epsilon_0.
\end{equation}
And a generous upper limit on the number of current sheets can be obtained
by taking $\ell'_s\sim \ell'_a$ (the distance an electron 
travels in order to get accelerated to LF $\gamma'_{max}$). Using
(\ref{la}) we find $\aleph_s \lae 5\times10^8 \zeta_B(R)\Gamma_2^{-5} 
R_{15}^{-1} L_{48}^{3/2}$. This upper limit is much too large to be of 
practical use. The length of an acceleration region can be much larger 
than $\ell'_a$ when electron acceleration is balanced by radiative
losses, and in that case far fewer number of PASs are needed to process 
magnetic energy to radiation.
Let us take $\ell'_s = \eta \ell'_a$, with $\eta\sim 10^3$ that is needed 
to ensure that the observed radiation is dominated by electrons in PASs 
(as opposed to electrons in inter-PAS regions)
and therefore the emergent spectrum is hard (see \S\ref{spectra}). This 
results in 
\begin{equation}
   \aleph_s \sim 5\times10^2 \zeta_B(R)\eta_3^{-2} \Gamma_2^{-5} R_{15}^{-1} 
             L_{48}^{3/2}.
\end{equation}

\section{Discussion}

This work was motivated in part by a puzzle regarding gamma-ray bursts.
A broad class of models for $\gamma$-ray emission from GRBs is based on 
the jet being baryonic, which moves with a Lorentz factor $\gae10^2$. The
kinetic energy of baryons in the jet is converted to particle thermal 
energy via a series of shocks, and radiated away via the synchrotron 
process. According to this model, the spectrum below the peak should 
be $f_\nu\propto \nu^{-1/2}$ or softer whereas the observed spectra 
for most bursts are close to $\nu^0$, i.e. much harder than the baryonic 
jet model predicts (e.g. Ghisellini et al., 2000; 
Kumar \& McMahon, 2008). The origin of this problem can be traced to 
the fact that particles are accelerated while crossing
the shock front but otherwise they cool rapidly as they travel down-stream.
Therefore, the number of electrons increases rapidly with decreasing LF 
($d n'_e/d\gamma'_e \propto \gamma'^{-2}_e$ or faster) and that is the 
reason for the soft spectrum for a generic model that is based on 
dissipation of baryonic jet energy in shocks.

What we find is that if the GRB jet were not baryonic but Poynting,
then the dissipation of magnetic fields and particle acceleration 
provides a way out this problem. This is because particles can be kept
in acceleration regions for a time much longer than the their
radiative cooling time, thereby preventing the development of 
a large population of lower energy electrons that give rise to a soft 
spectrum. The spectrum of radiation from electrons in the region in 
between PASs (where electrons are undergoing cooling without acceleration) 
is soft like that it is for the shock model, but the spectrum emanating 
from PASs is hard because the powerlaw index for the particle distribution 
function in current sheets has $p\approx 1$.
We have shown in \S\ref{spectra} that the observed spectrum, which is a
superposition of contributions from the two regions (PASs and inter-PASs),
is hard when the average time electrons spend in acceleration region
is much larger than their synchrotron cooling time. 

One of the general results reported here may be able to determine whether 
a jet is baryonic or Poynting --- for a Poynting jet, the spectrum below the 
peak softens with decreasing frequency, which is opposite to the case of a 
baryonic jet where shocks convert jet energy to radiation via the synchrotron 
process.

\section{Acknowledgments}

PK would like to thank Fan Guo for useful discussions. We are grateful to 
Sera Markoff and an anonymous referee for offering numerous suggestions 
that significantly improved the paper.

\end{document}